\documentclass[traditabstract]{aa}

\def\kms{km\,s$^{-1}$}

\usepackage{graphicx}
\usepackage{txfonts}
\usepackage{graphicx,epsfig,fancyhdr,rotating,amsmath,natbib}
\usepackage{natbib}
\usepackage{xcolor,ulem}

\begin{document}





\title{Transient and asymmetric dust structures in the TeV-bright nova RS Oph revealed by spectropolarimetry.}

   \author{Y. Nikolov\inst{1}\fnmsep\thanks{\email{ynikolov@nao-rozhen.org}},
   G. J. M. Luna\inst{2},
   K.A. Stoyanov\inst{1}, 
   G. Borisov\inst{1,3},          
   K. Mukai\inst{4}, 
   J. L. Sokoloski\inst{5} and
   A. Avramova-Boncheva\inst{1}
          	  }
   \institute{Institute of Astronomy and National Astronomical Observatory, Bulgarian Academy of Sciences, 
                 72 Tsarigradsko Chauss{\'e}e Blvd., BG-1784 Sofia, Bulgaria 
		 \and 
	      	 CONICET-Universidad Nacional de Hurlingham, Av. Gdor. Vergara 2222, Villa Tesei, Buenos Aires, Argentina 
		        \and
        Armagh Observatory and Planetarium, College Hill, BT61 9DG, Armagh, Northern Ireland, UK
        \and University of Maryland, Baltimore County, Baltmore, MD, USA
        \and Columbia Astrophysics Lab, 550 W 120th St, New York, NY, USA
         }

   \date{Received May 24, 2023; accepted  ...  .. 2023}

\abstract{A long-standing question related to nova eruptions is how these eruptions can lead to the formation of dust despite the ostensibly inhospitable environment for dust within the hot, irradiated ejecta. Novae in systems such as the symbiotic binary RS Ophiuchi (RS Oph), in which ejecta from the white dwarf collides with pre-existing circumstellar material fed by the wind from the red-giant companion, offers a particularly clear view of some nova shocks and any associated dust production. Here we use spectropolarimetric monitoring of the recurrent nova RS Oph starting two days after its eruption in 2021 August to show that: 1) dust was present in the RS Oph system as early as two days into the 2021 eruption; 2) the spatial distribution of this early dust was asymmetric, with components both aligned with and perpendicular to the orbital plane of the binary; 3) between two and nine days after the start of the eruption, this early dust was gradually destroyed; and 4) dust was again created, aligned roughly with the orbital plane of the binary, more than 80 days after the start of the outburst, most likely as a result of shocks that arose as the ejecta interacted with circumbinary material concentrated in the orbital plane. Modelling of X-rays and very-high energy GeV and TeV emission from RS Oph days to months into the 2021 eruption suggests that collisions between the ejecta and the circumbinary material may have led to shock formation in two distinct regions -- the polar regions perpendicular to the orbital plane where collimated outflows have been observed after prior eruptions, and a circumbinary torus in the orbital plane. The observations described here indicate that dust formed in approximately the same two regions, supporting the connection between shocks and dust in novae and revealing a very early onset of asymmetry. The spectropolarimetric signatures of RS Oph in the first week into the 2021 outburst indicate: 1) polarized flux across the $H_{\alpha}$ emission line and 2) position angle orientation relative to the radio axis are similar to the spectropolarimetric signatures of AGNs.}

   \keywords{Stars: binaries: symbiotic -- techniques: polarimetric   --  stars: individual: RS~Oph}

   \titlerunning{RS~Oph - asymmetry of the ejected material after the 2021 outburst.}
   \authorrunning{Nikolov et al.}
   \maketitle
%

\section{Introduction}

 With its seven recorded outbursts, RS~Oph is perhaps the best-studied recurrent nova. It belongs to a small group of objects in which a white dwarf (WD) accretes from the wind of its red giant companion and ignites under degenerate conditions with a sudden and major increase in luminosity when significant material has accumulated on its surface. The RS~Oph binary system consists of a massive carbon-oxygen WD with M$_{WD}$=1.2-1.4 $M_{\odot}$ \cite[where $M_{WD}$ is the mass of the WD, e.g.,][]{2017ApJ...847...99M} and a M2 III red giant \citep{2018MNRAS.480.1363Z} in a 453.6\,day orbit with an inclination of {\it{i}} = $49^{\circ}$--52$^{\circ}$ \citep{2009A&A...497..815B}.
 
 Detailed observations during the previous outburst allowed us to identify asymmetric structures on the expanding material \citep[e.g.,][]{2006Natur.442..279O, 2007ApJ...665L..63B,2009ApJ...703.1955R,2008ApJ...688..559R,2008ApJ...685L.137S, 2009ASSP...13..373O,2022ApJ...926..100M} 
 whose origin is still a matter of debate: Are the asymmetries generated once the expanding ejecta interacts with the surrounding circumstellar material?; 
 Are they the direct result of radio jets anchored in the accretion disk? 

 After the 2006 outburst, infrared observations taken as early as 3.8\,days after the start of the outburst indicated the presence of dust, mostly the silicate type, at about 17 AU from the WD, suggesting that this dust was already present before the outburst and survived the harsh environment during it \citep{2008ApJ...677.1253B}. Later in the outburst, dust was detected on days 208 after the 2006 eruption by \citet{2007ApJ...671L.157E}. \citet{2017MNRAS.469.1314D} conducted a theoretical study of the possibility that dust can be created behind radiative shocks during nova outbursts and concluded that under certain conditions, silicate-type dust can be created as early as one or two days into the outburst. This hypothesis can be tested during the 2021 outburst of RS~Oph.

The most recent outburst of RS~Oph, on 2021 August 8.93 UT, was first reported by K. Geary\footnote{http://www.cbat.eps.harvard.edu/iau/cbet/005000/CBET005013.txt}.
Hereafter we adopt the time of eruption determined by \citet{munari2021} as $T_0$=2459435.00 JD. We present spectropolarimetric observations of the recurrent nova RS~Oph from day 2 to day 352 after the start of the recent 2021 outburst, aimed at detecting variable intrinsic linear polarization and the presence of dust. In Section \ref{sec:obs} we present details of the observations and analysis. Sections \ref{sec:results}, \ref{sec:disc} and \ref{sec:concl} highlight our results, discussion and conclusions, respectively.

\section{Observations and Data Analysis \label{sec:obs}}

We obtained spectropolarimetric observations of RS~Oph on eleven nights, with nine of them after the 2021 outburst and two observations taken 26 and 16 months before, during quiescence. 

The spectropolarimetric observations were secured with the 2-Channel-Focal-Reducer Rozhen (FoReRo2)\footnote{It is based on data obtained with the Rozhen telescope (Bulgaria)}, similar to that described by \citet{2000KFNTS...3...13J}, attached to the Cassegrain focus of the 2.0m RCC telescope of the Bulgarian Rozhen National Astronomical Observatory. 
A rotatable Super-Achromatic True Zero-Order Waveplate~5 \footnote{http://astropribor.com/waveplates/} retarder (APSAW-5) was added to FoReRo2. The observations consist of sets of eight polarized spectra. The resolving power is $R \approx 1100$. A beam-swapping technique was used \citep{2009PASP..121..993B} to minimize the instrumental polarization. The instrumental polarization was corrected using a standard star with zero degrees of polarization. The offset between the position angle in the celestial and instrumental polarization was corrected using strongly polarized standard stars.\\
For the spectral observations we used two instruments - the FoReRo2 for the low-resolution observations and the ESPERO spectrograph for the high-resolution observations. The ESPERO spectrograph is a fibre-fed Echelle spectrograph mounted on the 2m telescope at Rozhen NAO 
\citep{2017BlgAJ..26...67B}. It has a resolution of $\sim$30~000 and covers the range 3900$\AA$ -- 9000$\AA$.\\
The journal of observations is given in Table \ref{tab.obsjournal}. The table contains the date (in format YYYY-MM-DD), UT, of the beginning of the observation, the number of days after the start of the 2021 outburst (the sign "-" indicates the days before the outburst), and the total exposure time.

\begin{table}
\centering
\caption{Journal of observations.}
\label{my-label}
\begin{tabular}{ r @{\hspace{0.5\tabcolsep}} r @{\hspace{0.5\tabcolsep}} r @{\hspace{0.5\tabcolsep}} r @{\hspace{0.5\tabcolsep}} r }
\hline
\hline
 \multicolumn{1}{c@{\hspace{0.5\tabcolsep}}}{Date} & \multicolumn{1}{c@{\hspace{0.5\tabcolsep}}}{UT-start} & \multicolumn{1}{c@{\hspace{0.5\tabcolsep}}}{Days after} & \multicolumn{1}{c@{\hspace{0.5\tabcolsep}}}{Exposure} \\
      &          & \multicolumn{1}{c@{\hspace{0.5\tabcolsep}}}{the outburst}       & \multicolumn{1}{c@{\hspace{0.5\tabcolsep}}}{[s]}      \\
\hline
\multicolumn{4} {c}{\bf{Spectropolarimetric observations with FoReRo2}}\\
\multicolumn{4} {l}{\bf{RS Oph}}\\
2019-07-02~&  ~21:48  &  $-768$	&   1440   \\
2020-04-19~&  ~23:26  &  $-476$	&   720    \\
2021-08-10~&  ~19:11  &  $2.0$		&   368   \\
2021-08-14~&  ~19:00  &  $5.9$		&   168    \\
2021-08-15~&  ~19:49  &  $6.7$		&   96     \\
2021-09-12~&  ~18:27  &  $34.8$		&   80    \\
2021-09-13~&  ~18:34  &  $35.8$		&   56    \\
2021-10-29~&  ~16:32  &  $81.8$		&   800   \\
2022-06-04~&  ~21:51  &  $299.9$	&   1920  \\
2022-06-05~&  ~22:15  &  $300.9$	&   1440  \\
2022-07-26~&  ~21:58  &  $352.0$	&   2160  \\
\multicolumn{4} {l}{\bf{2MASS~17500625-0643499}}\\
2023-07-17~&  ~21:46  &  $708$	&   2400  \\
\hline
\multicolumn{4} {c}{\bf{Spectral observations of RS Oph with ESPERO}}\\
2019-07-17~&  ~20:13  &  $-753$	&   3600  \\
2021-09-17~&  ~18:36  &  $40$	&   120   \\
2022-05-20~&  ~21:27  &  $285$	&   3600  \\

\hline
\hline
\label{tab.obsjournal}
\end{tabular} 
\\
 
\end{table}

\section{ Results \label{sec:results}}

\subsection{Spectral evolution}

\begin{figure}[htb]
    \begin{center}
      \includegraphics[width=0.5\textwidth, angle=0]{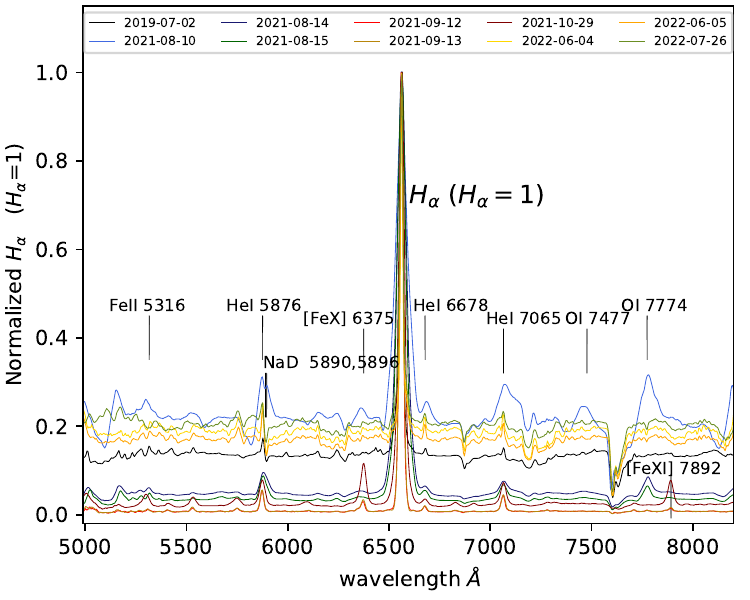}
     \end{center}
         \caption[]{Spectral evolution of RS~Oph after the 2021 outburst. For comparison, we also include a spectrum taken in quiescence, on July 2019. The most prominent lines are highlighted.} 
\label{fig.specevolution}
\end{figure}

On Figure~\ref{fig.specevolution} we present the spectral evolution of RS~Oph in the period August -- October 2021 based on the low-resolution data. 
The most prominent feature on the spectra is the H$\alpha$ emission line. After the outburst, the equivalent width of the H$\alpha$ line is increased, reaching $\sim$ $-$3000\,\AA\ on 2021 Sep, followed by a decrease. The complex structure of the line is not well defined due to the low resolution of the spectra. On 2021 Aug 10 -- 2 days after the outburst, a P~Cyg profile with velocity up to $\sim$ -4260~\kms is detected. The measured velocity is in agreement with the value, reported by \citet{2021ATel14852....1M}.\\
HeI emission lines at 5876~\AA, 6678~\AA\ and 7065~\AA\ rapidly evolve after the outburst, becoming prominent on the spectra after 2 days. The OI lines at 7477~\AA\ and 7773~\AA\ respectively are well-defined, especially on the spectrum obtained on 2021 Aug 10. 
The most prominent Fe feature in the spectra is the FeII multiplet at 5316~\AA, being more evident on the earliest spectrum. In addition, some coronal lines of [FeX] and [FeXI] at 6375~\AA\ and 7892~\AA\ respectively are visible on the later spectra, when RS~Oph entered a supersoft source phase \citep{2021ATel14894....1P, 2021ATel14895....1M}.

\begin{figure}[htb]
    \begin{center}
      \includegraphics[width=0.5\textwidth, angle=0]{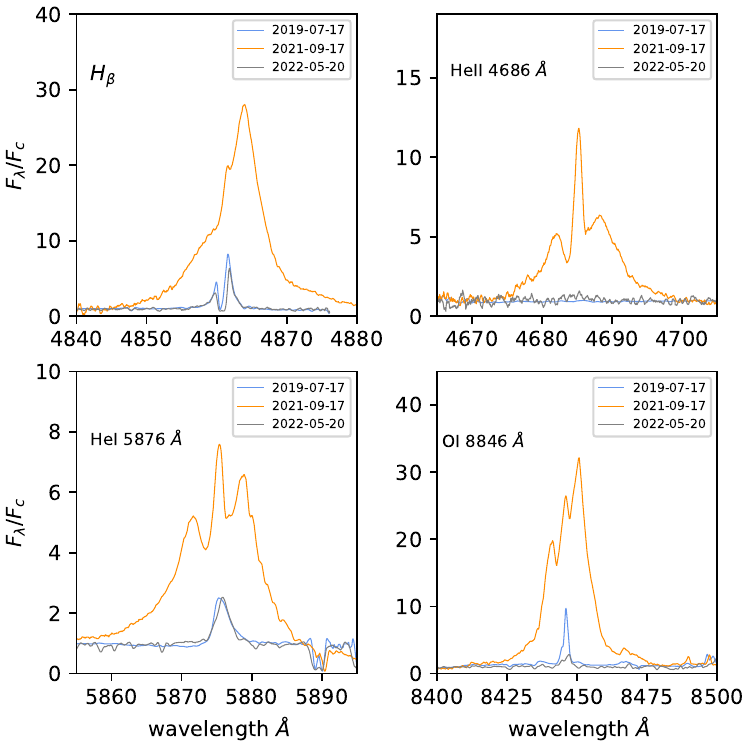}
     \end{center}
         \caption[]{Selected high-resolution emission lines in the spectra in quiescence (2019) and 40 and 285 days after the outburst. The line profiles are very complex, with at least three components at different radial velocities.} 
\label{fig.highresolutionspectra}
\end{figure}

On Figure~\ref{fig.highresolutionspectra} we present the spectral evolution of selected emission lines in the spectra of RS~Oph based on high-resolution data. The observations were obtained on 2021 September 17 (40 days after the outburst) and on 2022 May 20 (285 days after the outburst). For comparison, we present also a spectrum obtained before the outburst, on 2019 July 17. The Balmer H$\beta$ emission line observed on 2021 September 17 is complex, revealing a structure similar to that observed in H$\alpha$ line \citep{2022BlgAJ..37...24Z}. A sharp P~Cyg profile is visible at the top of a strong and broad emission.\\
The triple-peaked He I line has radial velocities of -235 km/s, -30 km/s and 130 km/s. The radial velocities of the peaks of the OI 8446 Å line are -190 km/s, -25 km/s and 137 km/s. The He~II lines are produced by shocked ejected material during its cooling phase, so the presence of the He~II~4686~\AA\ line 40 days after the outburst confirms the shock wave \citep{2023A&A...674A.139A}.

The spectral observations reveal a typical spectral behaviour of a nova with a complex structure dominated by broad Balmer, He, O, and Fe lines with some P~Cyg profiles, evolving towards a supersoft source phase.

\subsection{Interstellar polarization toward the Recurrent Nova RS~Oph}

\begin{figure}[htb]
    \begin{center}
      \includegraphics[width=0.47\textwidth, angle=0]{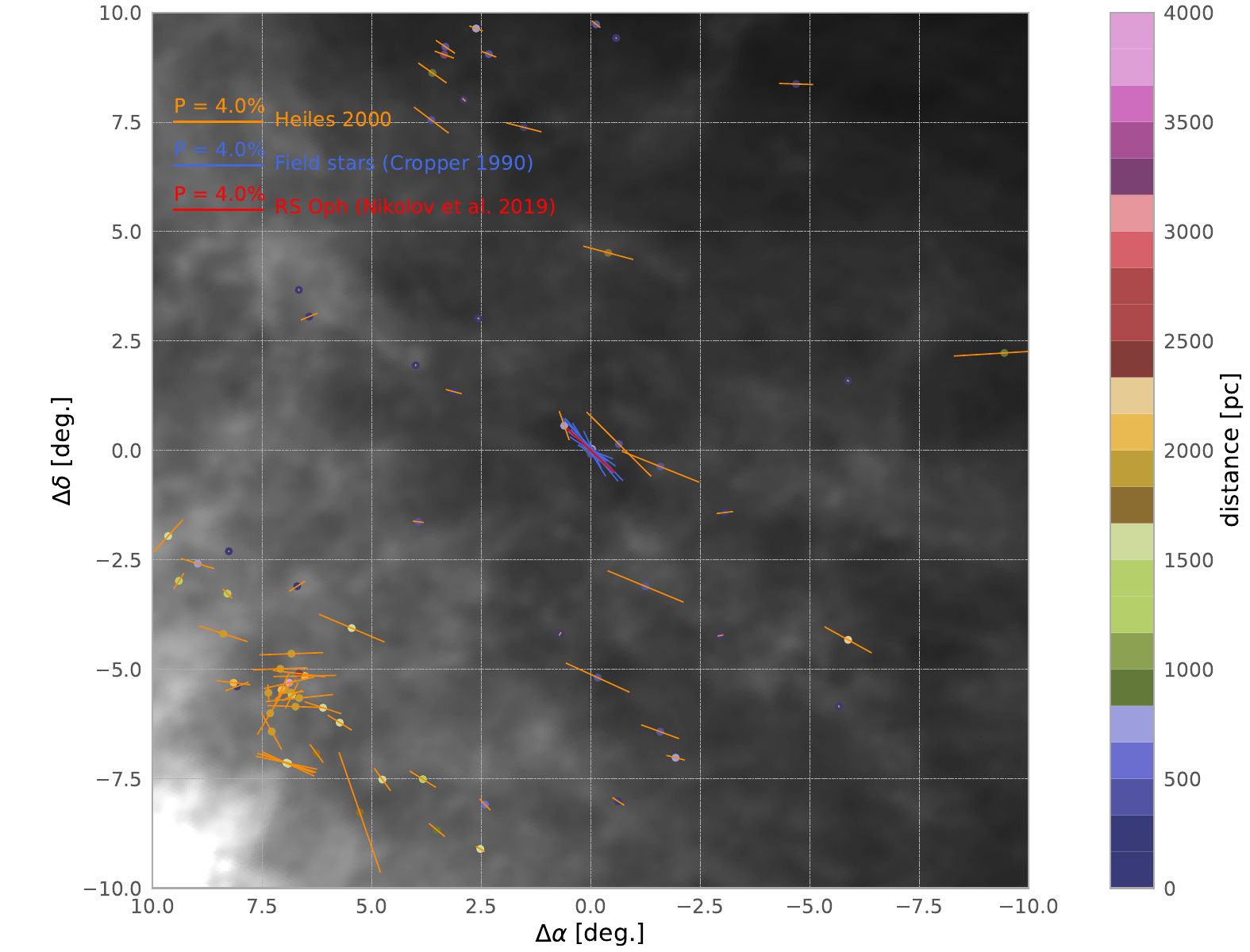}
     \end{center}
         \caption[]{The interstellar polarization of the field stars around RS~Oph \citep{2000AJ....119..923H, 1990MNRAS.243..144C}. The degree of polarization is proportional to the length of the orange straight lines. The horizontal bar at the top left presents 4.0\% polarization. The P.A. of the stars in the direction of RS~Oph (with blue bar) is similar to that of RS~Oph during quiescence \citep[red line;][]{2019AcA....69..361N}. The color of every star corresponds to its distance. The background image represents 100 $\mu$m dust emission maps \citep{1998ApJ...500..525S}.}
	 
\label{fig.3Dpol}
\end{figure}

The observed polarization is a vectorial sum of intrinsic polarization and interstellar polarization. The recurrent nova phenomenon gives us the opportunity to study an object before and after the nova outburst. A 3D map of the interstellar polarization in the field around RS~Oph is presented in Figure~\ref{fig.3Dpol}. The data are taken from \citet{2000AJ....119..923H}, \citet{1990MNRAS.243..144C}  and \citet{2019AcA....69..361N}. The observed degree of polarization and position angle of RS~Oph during quiescence is similar to the degree of polarization and position angle of the stars of the vicinity of RS~Oph published by \citet{1990MNRAS.243..144C}. Moreover, the spectropolarimetric observations during quiescence do not show the characteristics of an object with an intrinsic polarization, such as a depolarization effect across the $H_{\alpha}$ emission line, a variable degree of polarization and a variable position angle \citep{2019AcA....69..361N}. 

\begin{figure}[htb]
    \begin{center}
      \includegraphics[width=0.5\textwidth, angle=0]{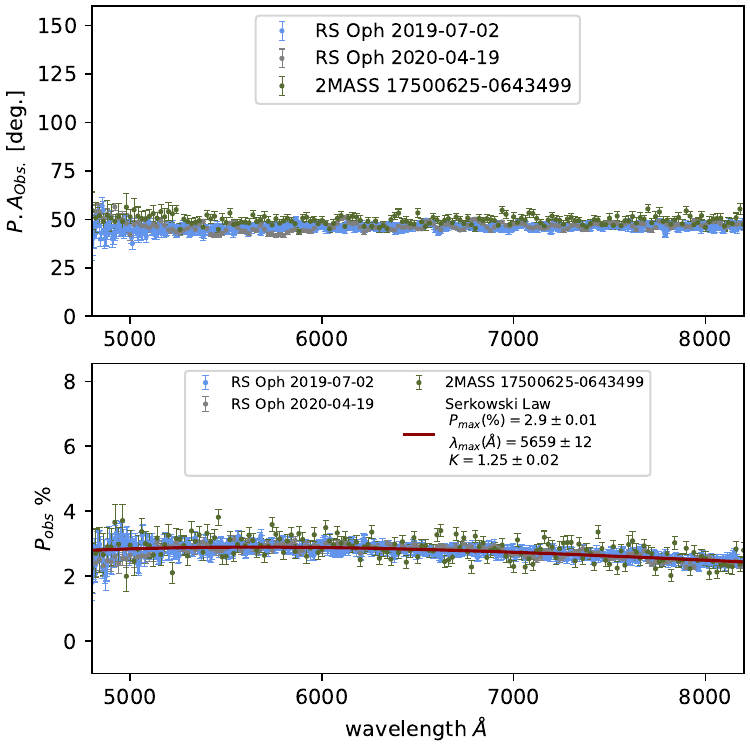}
     \end{center}
         \caption[]{Observed position angle (top) and degree of polarization (bottom) of RS Oph and 2MASS 17500625-0643499. The red curve represents fit with Serkowski’s law \citep{1975ApJ...196..261S}.}	 
\label{fig.2MASS}
\end{figure}

On Figure \ref{fig.2MASS} we show the observed degree of polarization and position angle of RS~Oph taken 26 and 16 months before the 2021 outburst, during quiescence, and position angle and degree of polarization of 2MASS~17500625-0643499, a nearby source at an angular distance of $\approx$ 2.2 arcmin. The wavelength dependence of the observed degree of polarization of 2MASS~17500625-0643499 is the same as the wavelength dependence of RS~Oph during quiescence at the same position angle.
The observations of RS~Oph during quiescence thus indicate that the observed pre-eruption polarization is interstellar, as no intrinsic polarization was detected \citep{2019AcA....69..361N,2017MNRAS.464.2784S}.

One of the methods to separate intrinsic and interstellar polarization is based on the depolarization effect of the emission lines.  Using this method, \citet{2006AJ....132..433K} determined the intrinsic polarization of Nova~V475~Scuti, in which a clear depolarization effect was well visible on day 25 after the outburst. Our observations demonstrate that in the case of RS~Oph, this method is applicable shortly after the outburst when the scattering region is relatively smaller than the emission line region and the degree of polarization in the $H\alpha$ emission line is approximately 0. The He I 5876 and $H\alpha$ lines have different degrees of polarization in the core of their profiles. Using them for the determination of the interstellar polarization would give a different (smaller) power law index. This method gives accurate results when the scattering region is much smaller than the emission-line region. In the case of RS~Oph, the scattering region gradually expanded into the emission-line region, and the depolarization effect decreased over time.

\subsection{Intrinsic polarization of RS~Oph after 2021 outburst}

To calculate the Stokes parameters of the intrinsic polarization $Q_{int}(\lambda)$ and $U_{int}(\lambda)$ of~RS~Oph, we derived these values from our observations and corrected them for the interstellar polarization using the equations:

\begin{equation}
Q_{int}(\lambda)=Q_{obs}(\lambda )-Q_{IS}(\lambda ), 
\end{equation}
\begin{equation}
U_{int}(\lambda)=U_{obs}(\lambda )-U_{IS}(\lambda ), 
\end{equation}
where $Q_{obs}(\lambda)$ and $U_{obs}(\lambda)$ are the Stokes parameters of the observed polarization after the 2021 outburst. $Q_{IS}(\lambda)$ and $U_{IS}(\lambda)$ represent the Stokes parameters of the interstellar polarization. As interstellar Stokes parameters, we used the parameters Q$_{IS}$($\lambda$) and U$_{IS}$($\lambda$) obtained on 2019 July 2. The degree of intrinsic polarization ($P_{int}(\lambda)$) and intrinsic position angle ($\theta_{int}(\lambda)$) were calculated using the following equations:

\begin{equation}
P_{int}(\lambda)=\sqrt{Q_{int}^2(\lambda) + U_{int}^2(\lambda)}, 
\end{equation}  

\begin{equation} 
  \theta_{int}(\lambda) = \frac{1}{2}~\arctan\frac{U_{int}(\lambda)}{Q_{int}(\lambda)} + \Theta_0,
\end{equation}
where $\Theta_0$ depends on the sign of $Q_{int}(\lambda)$ and $U_{int}(\lambda)$ \citep{2009PASP..121..993B}.

\begin{figure*}[htb]
    \begin{center}
      \includegraphics[width=\textwidth, angle=0]{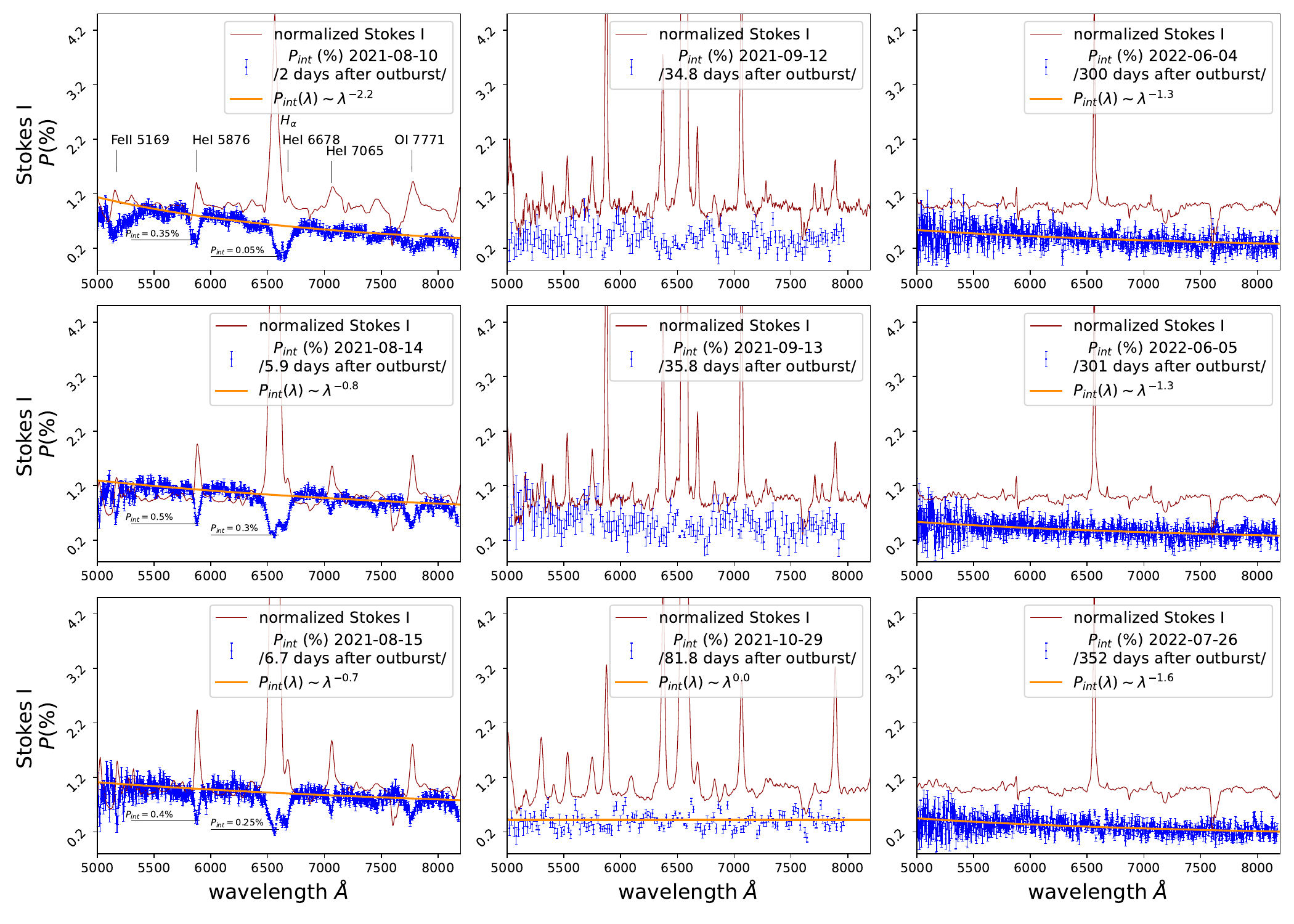}
     \end{center}
         \caption[]{Intrinsic polarization of RS~Oph after the 2021 outburst. The degree of polarization is presented in blue. Strong depolarization effects are well visible in the first week after the start of the 2021 outburst (left column). The normalized intensity is presented in red. The orange line represents the power law fit of the degree of polarization data. The index of the power law increases in the first week after the start of the 2021 outburst (left column). Later, from day 300 to day 352, the power law index decreases (right column).}
\label{fig.Pint}
\end{figure*}

In Figure \ref{fig.Pint} we  present the normalized intensity (normalized Stokes I) parameter and the degree of intrinsic polarization as a function of wavelength for each of the spectra taken during our campaign. The polarization spectra on days 34.8 (2021 September 12) and 35.8 (2021 September 13) are noisy because of the low continuum signal. The spectra obtained on days 34.8, 35.8 and 81.8 after the 2021 outburst are thus binned with a dispersion of 20\,\AA/pix. The maximum degree of intrinsic polarization, of about 1\%, is observed on day 5.9.

Strong depolarization effects are observed in FeII 5169\,\AA, HeI 5876\,\AA, ~$H\alpha$, HeI 6678\,\AA, HeI 7065\,\AA\ and OI 7771\,\AA\ emission lines until 6.7\,days after the outburst. 
The small polarization of the $H\alpha$ and HeI 5876\,\AA\ emission lines on days 2, 5.9, and 6.7 is noticeable. The polarization in the $H\alpha$ emission line is 
approximately 0.05\% on day 2 and 0.3\% on days 5.9 and 6.7. The HeI 5876\,\AA\ emission line polarization is approximately 0.35\% on day 2 and 0.5\%--0.4\% on days 5.9 and 6.7, respectively. The typical error 
on the degree of polarization in the emission lines is 0.05\%. The depolarization in the emission lines suggests that the scattering material is located within the emission line region. On the other hand, the polarization in the HeI 5876\,\AA\ emission line is higher than the $H\alpha$ emission line. This is because the $H\alpha$ region is more extended than HeI 5876\,\AA\ region. It is worth noting that the depolarization effect decreased with time. This can be explained by the scattering region expanding into the emission-line region.

\begin{figure}[htb]
    \begin{center}
      \includegraphics[width=0.47\textwidth, angle=0]{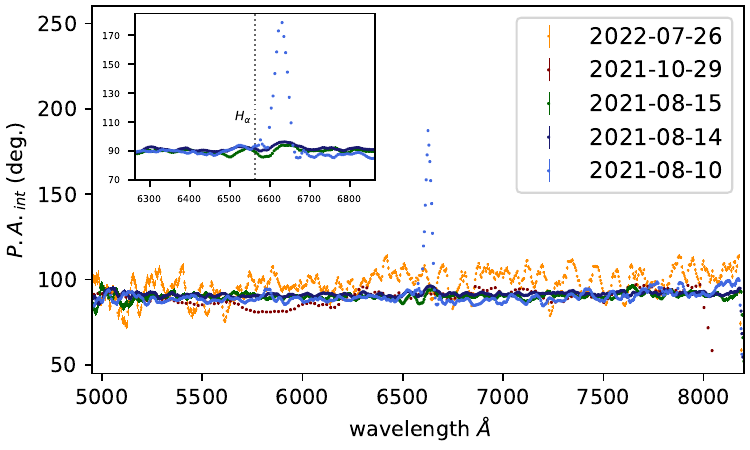}     
      \end{center}
         \caption[]{Intrinsic position angle of RS~Oph after the 2021 outburst. Aside from the 180$^{\circ}$ position angle around the H$\alpha$ region on day 2, the position angle of the intrinsic polarization is on average 90$^{\circ}$, implying that the scattering region that produced the observed polarization was oriented perpendicular to the elongated structures observed at other wavelengths during the last two outbursts. The data were smoothed with a moving average over seven points to better illustrate the position angle structures.}
	 
\label{fig.pa.int}
\end{figure}

Figure \ref{fig.pa.int} shows the position angle  of the intrinsic polarization as a function of wavelength from the observations taken after the 2021 outburst (days 2, 5.9, 6.7, 81.8 and 352). We omit data on days 34.8, 35.8, 300 and 301 due to their low signal-to-noise ratios. The data presented in Figure \ref{fig.pa.int} were smoothed using a moving average of over seven points to better illustrate the position angle structures across the $H_{\alpha}$ emission line.
The position angle of the intrinsic polarization is approximately $90^{\circ}$ throughout the wavelength range explored, with the exception of the region around H$\alpha$ on day 2, which shows a position angle of 180$^{\circ}$ (see discussion in Section \ref{sec:evol.pol}). The 90$^{\circ}$ position angle, which implies a scattering structure that extends predominantly in the N-S direction on the sky, is aligned with the highly collimated outflows detected during the outbursts in 2006 \citep[e.g.][]{2008ApJ...688..559R} and 2021 \citep{2022A&A...666L...6M}.

\subsubsection{Wavelength dependence of the intrinsic polarization}
 
\begin{figure}[htb]
    \begin{center}
      \includegraphics[width=0.47\textwidth, angle=0]{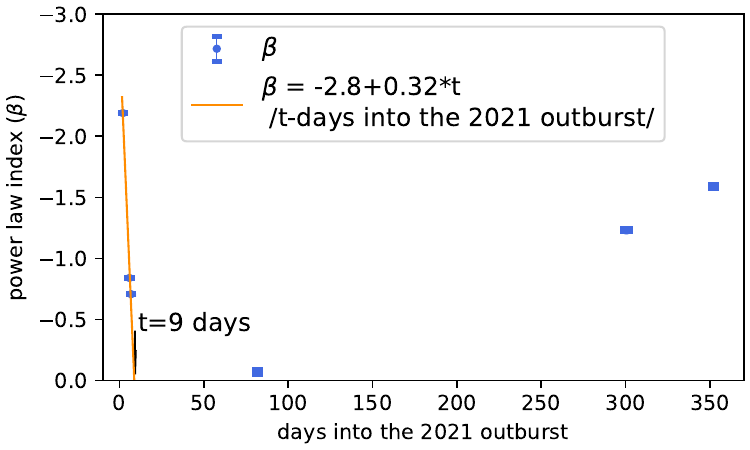}
     \end{center}
         \caption[]{Long-term variability of the index of the power law fit, $\beta$, to the polarization degree dependence with wavelength.}
	 
\label{power.law.index}
\end{figure}

The degree of intrinsic polarization after the 2021 outburst increases toward 
shorter wavelengths. We use a monotonic function (power-law) as a description of the variation of the degree of intrinsic polarization with the wavelength.
We fit the continuum data with a power law as:
 
\begin{equation}
P_{int}(\lambda)=c\lambda^{\beta}, 
\end{equation} 
where $\rm{c}$ is a scaling constant and $\beta$ is a power law index. This index indicates the nature of the scattering process: $i$) Mie scattering by dust grains or, $ii$) Thompson scattering by electrons. The first process is wavelength dependent whereas Thompson scattering is not. Figure \ref{power.law.index} shows the power-law index as a function of time after the outburst. The index increases from $\beta$ = $-$2.2 on day 2 and reaches  $\beta$ = 0,  which corresponds to Thompson scattering, on day 9. 
After that, the index decreased again until day 352, reaching values of $\beta$ = $-$1.6. 

\subsubsection{Long-term variability of the degree of intrinsic polarization and position angle in broad-bands}

We used the IRAF SBAND procedure to calculate broadband linear polarization. Two sets of bands were chosen -- standard V band and filter C. We selected a region with a weak depolarization effect in the wavelength range between $\lambda_{1} = 5200\AA$ and $\lambda_{2} = 5600\AA$ to define the filter C. The transmission function of the C filter is 100\,\%.

\begin{figure}[htb]
    \begin{center}
      \includegraphics[width=0.47\textwidth, angle=0]{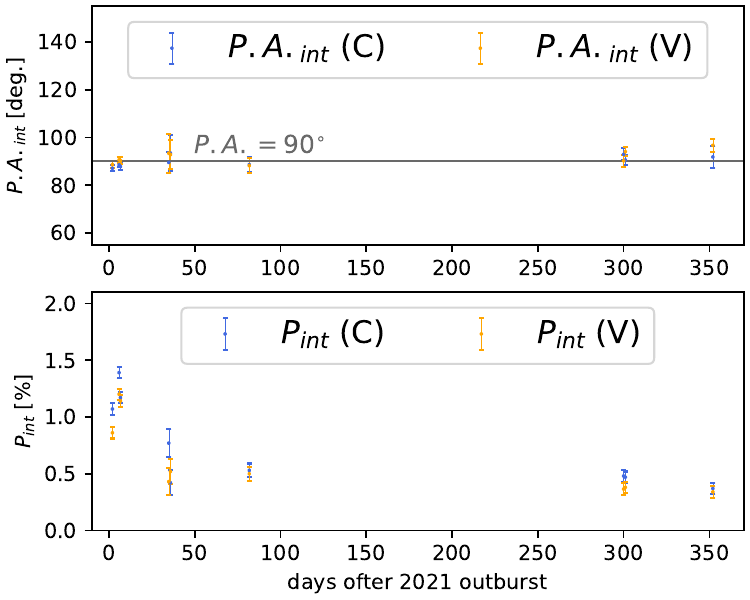}
     \end{center}
         \caption[]{Long-term variability of the position angle (top) and degree of intrinsic polarization (bottom) of RS~Oph after the start of the 2021 outburst. The P.A. was practically constant throughout our observing campaign, while the degree of intrinsic polarization reached a maximum at day 5.9 and has been decreasing since then.}
	 
\label{fig.JDPint}
\end{figure}

Long-term variability of the degree of intrinsic polarization and position angle are presented in Figure \ref{fig.JDPint}. $P_{int}$(V) and $P_{int}$(C) represent the degree of intrinsic polarization in the V band and in the C band, respectively. The intrinsic polarization increased from day 2 to day 5.9 after the outburst. After day 5.9 the degree of intrinsic polarization decreased, and we obtain 
$P_{int} (V) = 0.34\% \pm 0.05\%$ and $P_{int} (C) = 0.37\%\pm 0.05\%$ on day 352 after the outburst. 
On day 5.9, we detected the greatest degree of intrinsic polarization, with $P_{int} (V) = 1.2\% \pm 0.05\%$ and $P_{int}(C) = 1.39\%\pm 0.05\%$. The position angle of the intrinsic polarization ($P.A._{int}(V)$ and $P.A._{int}(C)$) was practically constant during the observations and had a value of approximately $90^{\circ}$. 

\subsubsection{Evolution of the polarized flux. }
\label{sec:evol.pol}

\begin{figure*}[ht!]
    \begin{center}
      \includegraphics[width=\textwidth, angle=0]{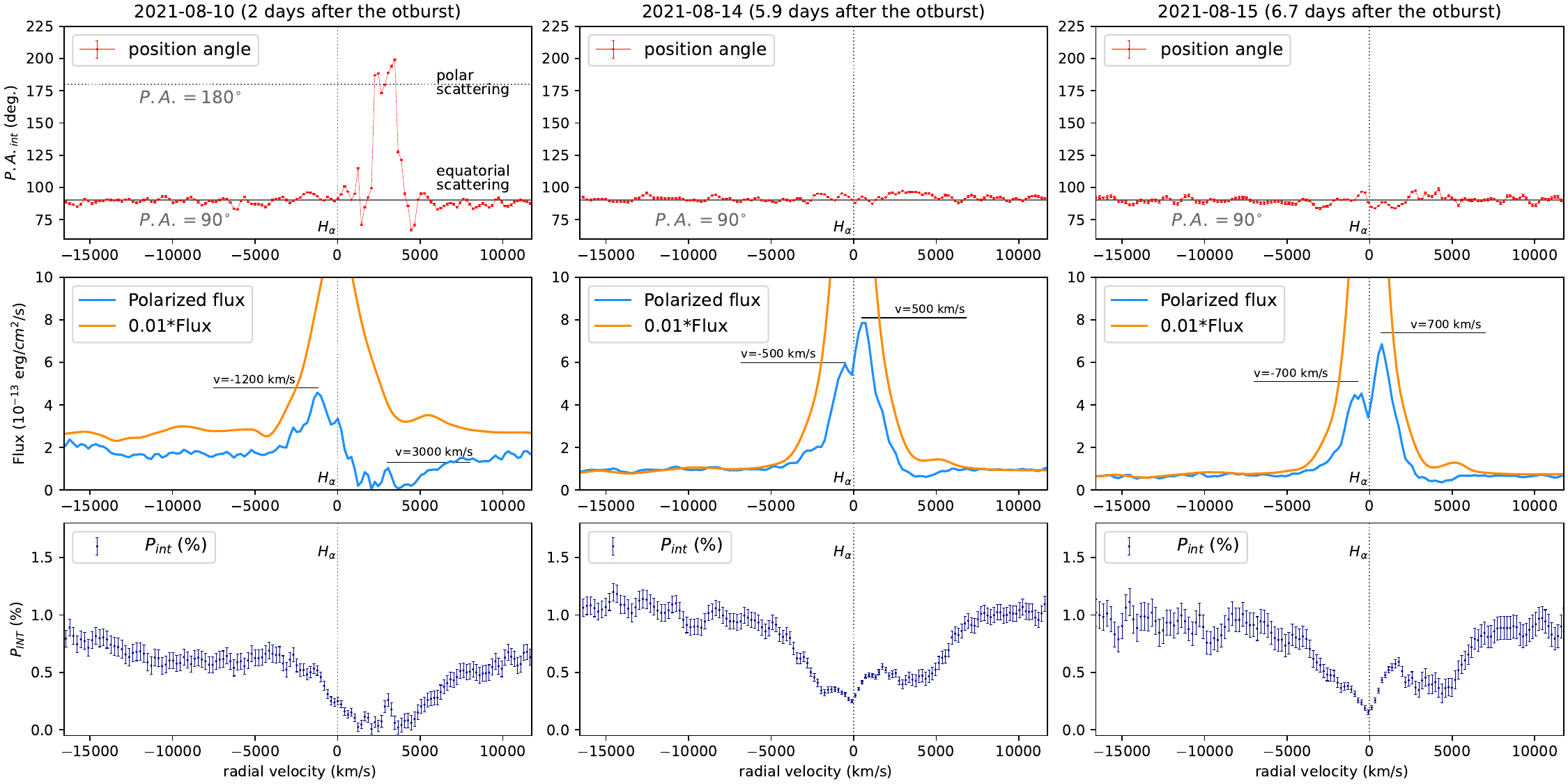}
     \end{center}
         \caption[]{Evolution of the polarized flux across the $H_{\alpha}$ emission line. The position angles (P.A.) are presented in red in the first row from day 2 (left) to day 6.7 (right) into the 2021 outburst. The P.A. on the 2nd day into the outburst (left column, top) shows a complex structure. The evolution of the polarized flux (in blue) and 0.01*flux (in orange) are presented in the second row. Asymmetries in the blue- and red-shifted peaks of the polarized flux are well visible. In the third row, the degree of polarization is presented.}
	 
\label{fig.polflux.evolution}
\end{figure*}

In Figure \ref{fig.polflux.evolution}, we present the position angle, the evolution of the polarized flux, and the degree of the intrinsic polarization during the first week of the outburst. 
The position angle (P.A.) of the intrinsic polarization on the 2nd day into the outburst (left column, top) shows a complex structure, with $P.A.\approx 90^{\circ}$ at wavelengths between $\lambda=6200$\,\AA\ and $\lambda=6600$\,\AA\ and $P.A.\approx 180^{\circ}$ for wavelengths between $\lambda=6600$\,\AA\ and $\lambda=6650$\,\AA. Radio interferometric imaging of RS Oph obtained 34\, days after the start of the 2021 nova outburst shows a bipolar structure \citep{2022A&A...666L...6M} in the east-west direction; the orbital plane is perpendicular to these radio lobes. When the $P.A.$ was $90^{\circ}$, it was thus 
aligned with the observed bipolar structure. When 
the $P.A.$ was $180^{\circ}$, it was aligned with the orbital plane. 
The wavelength range with $P.A.=180^{\circ}$ corresponds to the small peak in the polarized flux spectrum near a velocity of $v=3000$\,\kms (see the left column in Figure \ref{fig.polflux.evolution}). The two main observed values of the position angle correspond to the mutually perpendicular planes.


\section{Discussion \label{sec:disc}}

\subsection{Quick dust formation after the outburst}

Our observations show the presence of dust as early as 2\,days after the outburst. The wavelength dependence of the intrinsic polarization gives information about the nature of the scattering process. If the scattering is dominated by dust, we expect an increase in the intrinsic polarization toward shorter wavelengths. By modelling the continuum of the polarized spectra, we found that it was strongly wavelength-dependent from day 2 until day 9 after the outburst. This finding provides the most direct observational support
of the presence of dust in a nova at such early times.

Although other novae have shown optical polarization within a few days of the start of the eruption, the case for early dust associated with the nova has been less definitive.
As early as 1.36\,days after the outburst of T~Pyx, 
\citet{pavana19} detected polarization that 
could be attributed to asymmetries during the outburst 
or the presence of silicate dust that survived the eruption; however, given the unknown amount of interstellar polarization toward T~Pyx, it was not possible for those authors to determine the wavelength dependence of the intrinsic polarization and how the intrinsic polarization and position angle varied with time, precluding a definitive conclusion about its origin. \citet{2000ApJ...540..429K} observed Nova V4444 Sgr from days 2 to 10 after the start of its 1999 outburst and found a wavelength-dependent intrinsic polarization with a power law index in the range: $-2 \lesssim \beta \lesssim -1$. Given that this dependence did not change with time, however, the authors concluded that the most likely origin was scattering by pre-existing dust located at several dozen AU from the central star. Nova~V1494~Aql was observed between days 2.39 and 3.39 after the start of its outburst by \citet{2001ApJ...552..782K}. In their Fig.~4, the visual negative slope of the degree of intrinsic polarization with the wavelength increases from day 2.39 to day 3.39 and decreases since then.

Evidence for dust formation in novae has been found through observations in the IR and optical at a much later phase of the outburst. Infrared observations of RS~Oph with the {\it Spitzer} telescope from 208 to 430\,days after the start of the 2006 eruption of RS~Oph revealed the presence of silicate dust in the environs of the system \citep[see][]{2007ApJ...671L.157E, 2008ASPC..401..260W}. \citet{2002A&A...384..504E} found that the wavelength-dependence of polarization in the case of V4362~Sgr suggests scattering by small dust grains (80\,days after the start of the outburst). In the case of V339~Del, \citet{2019ApJ...872..120K} used the continuum from the neighbouring emission-free regions as values for the interstellar polarization. Nova V339~Del produced dust grains around 35\,days into the outburst \citep{2017MNRAS.466.4221E}.  


Our detection of wavelength-dependent intrinsic polarization indicates that, most likely, 
dust particles were created in dense regions shielded from the strong radiation field of the eruption. Such wavelength-dependent polarization
was not observed during quiescence (down to our detection error of 0.05\%), which strongly suggests that dust was not present during the quiescence phase.
Moreover, Cropper (1990) found that the observed polarization of RS~Oph a year after the start of the 1985 outburst was comparable with the polarization of 
nearby stars. 

The possibility that dust can be created in radiative shock regions was studied by \citet{2017MNRAS.469.1314D}, who concluded that dust can be formed in, dense, shielded shock regions. 
These dense shocks, however, have been modelled by \citet{2014MNRAS.442..713M,2015MNRAS.450.2739M} in the context of novae in CVs; similar studies tailored for embedded nova such as RS~Oph are necessary to establish if dense regions behind 
shocks in those systems are also expected to be conducive to dust formation. Our data show that dust was present between days 2 and 9, and at about this same time, the $\gamma$-ray emission 
detected by Fermi peaked. The GeV-TeV emission has been modelled as due to multiple shocks, with the "slow" component first expanding into a high-density environment \citep{2022arXiv221102059D}. Our finding thus provides strong observational support 
that the scenario proposed by \citet{2017MNRAS.469.1314D} is plausible for embedded novae as well as classical novae \citep[see, e.g.,][for a discussion of dust formation in classical novae]{chomiuk2021, babul2022}.

The changes in the wavelength dependence of the intrinsic polarization with time, parameterized by the power law index $\beta$, indicate that the process of scattering switched from a mix of dust with different-sized grains to Thompson scattering after day 9. This switch indicates that any dust that was formed within the first week or so of the eruption quickly evaporated. Conditions then became favourable for dust formation again much later in the outburst.
In their infrared observation on day 12 after the 2021 outburst, \citet{2021ATel14866....1W} did not find evidence of dust. The spectropolarimetric observations on days 300 and 352 after the outburst do show the presence of dust grains. Further infrared observations can confirm the presence of dust months to years after the 2021 outburst.

\subsection{Early development of asymmetries in the ejecta}

\begin{figure*}[ht!]
    \begin{center}
      \includegraphics[width=0.97\textwidth, angle=0]{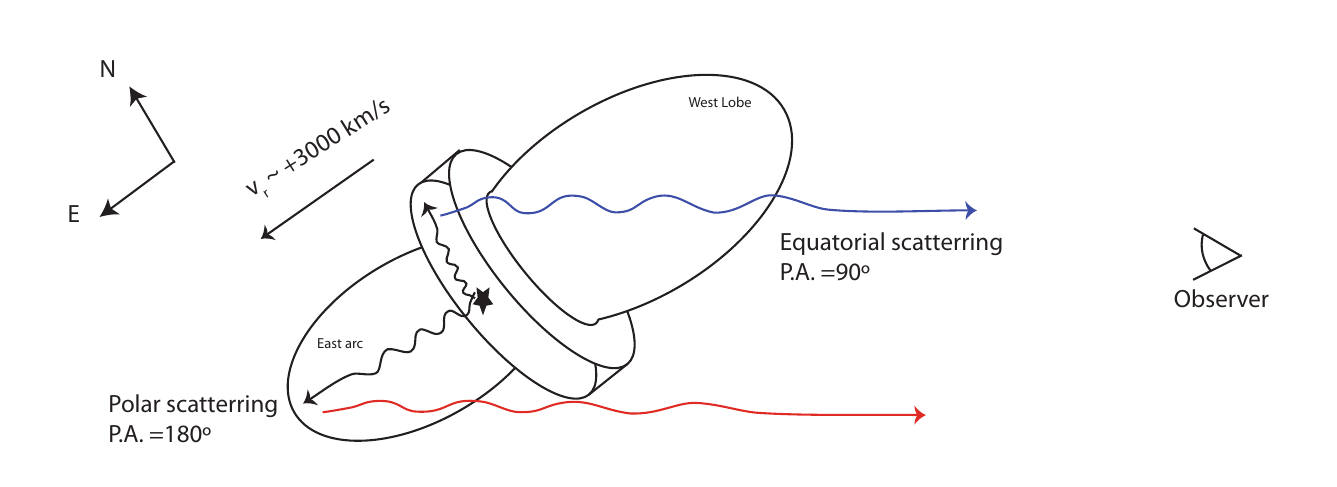}     
      \end{center}
         \caption[]{Schematic view of the polar and equatorial scattering regions that are detected at different position angles and their contribution to the polarized flux. }
	 
\label{fig.cartoon}
\end{figure*}

After the previous outburst in 2006, observations at a variety of wavelengths revealed evidence of the presence of asymmetric structures aligned preferentially in the East-West direction. \citet{2007A&A...464..119C} found a structure moving 
rapidly in the E-W direction 5.5\,days into the outburst. \citet{2007ApJ...665L..63B} and \citet{2009ApJ...703.1955R} found also E-W extended emission in the HST images taken 155 and 449\,days after the start of the 2006 outburst. Radio observations reported by \cite{2006Natur.442..279O} two weeks into the outburst, \cite{2008ApJ...688..559R} 20.8 and 26.8\,days into the outburst, and \cite{2008ApJ...685L.137S} 5 and 8 weeks into the outburst, reported extended emission in the E-W direction as well. X-ray observations reported by \citet{2022ApJ...926..100M} also show evidence of E-W asymmetries 1254 and 1927\,days after the start of the outburst. 
Radio interferometric imaging of RS Oph 34 days into the 2021 outburst shows a bipolar structure with two components: east arc (EA) and west lobe (WL), expanding perpendicular to the orbital plane in opposite directions \citep{2022A&A...666L...6M}.
All observations of asymmetry indicate a bipolar structure with a dominant East-West orientation. 

Our detection of linear polarization on day 2 constitutes the earliest evidence of an asymmetric structure in RS~Oph reported 
to date. With a position angle of 90$^{\circ}$ 
aligned with the previously known E-W structures, our finding indicates that some structure was present in the equatorial plane before day 2, consistent with the scenario 
in which the ejecta is shaped after its interaction with an equatorial density enhancement \citep{2022A&A...666L...6M}. From the orientation of the radio lobes, which are perpendicular to the orbital plane, we can conclude that $P.A.=90^{\circ}$ corresponds to the scattering of light by material in the orbital plane. On the other hand, $P.A.=180^{\circ}$ corresponds to the scattering of light by material in the polar regions.

The polarized flux represents the spectrum of the scattered light. We observed a noticeable asymmetry of the polarized flux on day 2 
-- a strong blue-shifted peak with a radial velocity of $v_{r}\approx-1200~km/s$ at a position angle $P.A.\approx90^{\circ}$ and a small red-shifted peak with a radial velocity of $v_{r}\approx +3000 km/s$ at a position angle $P.A.\approx180^{\circ}$ (Figure \ref{fig.polflux.evolution}). The interaction between the ejecta and an equatorial density enhancement is responsible for the formation of a ring-like structure in the orbital plane \citep[see][]{2022A&A...666L...6M}. As depicted in Figure \ref{fig.cartoon}, the west lobe is moving toward us and the portion of the torus that is moving away from us (red-shifted) is hidden by the west lobe. The material within the west lobe absorbs the scattered light from the red-shifted part of the torus. This half-torus structure is similar to the structure observed in radio wavelengths by \citet{2006Natur.442..279O} after the 2006 outburst. It is possible to separate the equatorial from the polar scattering regions only in the case when the red-shifted component of the equatorial torus is hidden. Such asymmetry of the polarized flux and position angle is not observed during the next observations between days 5.9 and 6.7. The P.A. on days 5.9 and 6.7 shows that matter in the orbital plane is now responsible for the scattering.
Further proof of the torus-like structure in the orbital plane is a triple-peaked profile of the permitted lines \citep{2022A&A...666L...6M}. 
The velocities of the double peak of the polarized flux across the $H_{\alpha}$ emission line are highlighted in Figure \ref{fig.polflux.evolution} and represent the radial velocities of the slowly expanding torus. Our values of the radial velocities of the torus and the east arc are consistent with the observations, presented by \citet{2007A&A...464..119C} 5.5 days after the 2006 eruption. \citet{2007A&A...464..119C} identiﬁed a slowly moving equatorial ring ($v_{r}\leq1800~km/s$) and a faster moving East-West bipolar ﬂow ($v_{r}\sim2500-3000~km/s$).

Similar torus-like structures have been observed in classical novae, but in this case, they are thought to arise from the propeller effect of the orbital motion on the ejecta \citep{2014Natur.514..339C}. Given the long orbital periods of symbiotic recurrent novae, it is not entirely clear if this effect could be at work or if the torus-like structure had been present in the system before the outburst.

\subsection{Different types of objects - similar spectropolarimetric signatures}

The similar spectropolarimetric signatures between RS Oph and AGNs relate to the following characteristics:

\begin{itemize}
    \item Polarized flux structure across the $H_{\alpha}$ emission line.
    \item Position angle orientation relative to the radio axis.
\end{itemize} 

Three-quarters of the AGNs with detected intrinsic polarization presented by \citet{2002MNRAS.335..773S} exhibit a decrease in the degree of polarization across the core of the $H_{\alpha}$ emission line. In two of them, Mrk 6 and Mrk 335, the red-shifted peak of the polarized flux is higher than the blue-shifted peak, similar to the polarized flux in RS Oph observed on day 5.9 and day 6.9 into the 2021 outburst (Fig. \ref{fig.polflux.evolution}). In the case of RS Oph, the depolarization in the emission lines is due to the scattering material being located within the emission line region. A similar explanation was suggested by \citet{1999AJ....118.1963C}. Those authors observed weakly polarized narrow lines in a sample of 13 FR~II radio galaxies. The narrow lines arise outside of the scattering region and the net effect would be a decrease of the degree of polarization in the core of the emission lines. A more complex explanation of the decrease in the degree of polarization for AGNs has been proposed by \citet{2002MNRAS.335..773S}, where the broad-line region 
takes the form of a rotating disc surrounded by a co-planar ring of scatterers -- this geometry will produce a high degree of polarization in the wings of the $H_{\alpha}$ emission line \citep[see][Sec~4.3]{2002MNRAS.335..773S}.
A similar explanation may also be valid for RS Oph if the accretion disc survives the outburst and lies in the same plane as the scattering region.

A position angle parallel to the radio axis (with $P.A.=90^{\circ}$ in the case of RS Oph) is similar to that observed in type 1 Seyfert galaxies, where the optical polarization P.A. is often parallel to the radio axis \citep{2004MNRAS.350..140S}, whereas $P.A.=180^{\circ}$ is most similar to type 2 Seyfert galaxies, 
in which the optical polarization position angle is perpendicular to the projected radio source axis \cite[e.g.,][]{1983Natur.303..158A,1990MNRAS.244..577B}.  
Scattering in a plane perpendicular to the radio axis produces a position angle parallel to the radio axis. On the other hand, scattering in a plane aligned with the radio axis would produce a position angle perpendicular to the radio axis \citep{2002MNRAS.335..773S}.
We suggest that similar geometry, torus and polar scattering region can reproduce similar spectropolarimetric signatures.

Theoretical results based on hydrodynamical simulations for classical novae predict that much of the outflowing mass is focused into the equatorial plane and forms spiral arms \citep{2022ApJ...938...31S}. Moreover, 3D hydrodynamical simulations of RS Oph in the quiescent phase show that the circumstellar mass distribution is highly structured with a mass enhancement in the orbital plane \citep{2008A&A...484L...9W,2016MNRAS.457..822B}. For RS~Oph, the interaction between the nova ejecta with the accretion disc and focused mass in the orbital plane forms an equatorial ring-like structure and limits the ejecta within a bipolar direction perpendicular to the orbital plane \citep{2016MNRAS.457..822B}.

For those novae that form an equatorial-like structure (or torus) during the outburst, the scattering geometry limits the orientation of the position angle, parallel or perpendicular to the torus, depending on the relative contribution of the equatorial and polar scattering region.

\section{Conclusions \label{sec:concl}}

The optical linear spectropolarimetric observations in the range from 5000\,\AA\ to 8200\,\AA\ of the recurrent nova RS~Oph after the 2021 outburst indicate that:

\begin{itemize}
	
     \item The variable degree of intrinsic polarization in RS~Oph with time and wavelength suggests the presence of dust as early as day 2. This dust could have been created in the strongly shielded, dense shock regions \citep{2007ApJ...671L.157E}.
     \item The continuum intrinsic polarization can be fitted with a power law. The power law index $\beta$ increased from $\beta$ = $-$2.2 on day 2 to $\beta$ = 0 on day 9. The flat polarization continuum suggests that dust was destroyed, and the scattering was then due to Thompson electron scattering. 
     In the last three observations, after day 300 into the 2021 outburst, the power index $\beta$ decreases to $\beta$=-1.6. This is an indicator of dust formation and the position angle $P.A.=90^{\circ}$ means that the dust is located in the orbital plane.     
     \item The polarized light two days after the outburst shows two position angles at $90^{\circ}$ and $180^{\circ}$ degrees. The position angle $P.A._{int} \approx 90^{\circ}$ is aligned with the radio/optical/X-rays elongated structures detected at much later times after the 2006 outburst. The 
     $P.A._{int} \approx 180^{\circ}$ degrees indicates that during these early times, there is a region responsible for the light scattering that is perpendicular to the ejecta in the polar directions. 
    These provide the first evidence of the development of asymmetries in the ejecta.
         
    \item We suggest that similar geometry (torus and polar scattering region) presented in AGNs and RS Oph, can reproduce similar spectropolarimetric signatures.
    
\end{itemize}

\begin{acknowledgements}
Y.N. acknowledges partial support by grants: K$\Pi$-06-M58/1 "Spectral and spectropolarimetric characteristics of the interstellar medium" and K$\Pi$-06-H28/2  "Binary stars with compact object" by the Bulgarian National Science Fund. K.S. acknowledges support by the Bulgarian National Science Fund -- project KP-06-Russia/2-2020, "An investigation of binary systems with a compact object using Doppler tomography". A.A.-B. acknowledges support by the Bulgarian National Science Fund -- project K$\Pi$-06-H58/3 "Environments of exoplanets: the role of coronal mass ejections and superflares". GJML is a member of the CIC-CONICET (Argentina) and acknowledges support from grant ANPCYT-PICT 0901/2017. J.L.S. acknowledges support from US NSF award AST-1816100. \\
The authors would like to thank for the fruitful discussion, comments, and support of Valeri Golev, Radoslav Zamanov, Luba Slavcheva-Mihova, Boyko Mihov, and Claudia Vilega Rodrigues. 
 
\end{acknowledgements}

\bibliographystyle{aa}
\bibliography{ref2.bib}

\newpage

\end{document}